\documentclass[conference]{IEEEtran}
\usepackage{caption} 
\usepackage{subcaption}
\usepackage{xcolor}
\usepackage{ifthen}
\usepackage{graphicx}
\usepackage{pdfpages}
\usepackage{enumitem}
\usepackage{listings}
\usepackage{amssymb}
\usepackage{amsmath}
\usepackage[hidelinks]{hyperref}
\usepackage[section]{placeins}
\usepackage{pdflscape}
\usepackage{color}
\usepackage{multirow}
\usepackage{multicol}
\usepackage{adjustbox}
\usepackage{tabularx}
\usepackage{amsmath}
\usepackage{float}
\usepackage{tcolorbox}
\usepackage[normalem]{ulem}
\usepackage{booktabs}
\usepackage{float}

\begin{document}
\title{Typhon:~Automatic Recommendation of Relevant Code Cells in Jupyter Notebooks}

\author{\IEEEauthorblockN{Chaiyong Ragkhitwetsagul, Veerakit Prasertpol, Natanon Ritta, Paphon Sae-Wong,\\ Thanapon Noraset and Morakot Choetkiertikul}
\IEEEauthorblockA{
\textit{Faculty of Information and Communication Technology, Mahidol University}, Nakhon Pathom, Thailand}
}

\maketitle

\begin{abstract}
At present, code recommendation tools have gained greater importance to many software developers in various areas of expertise. Having code recommendation tools has enabled better productivity and performance in developing the code in software and made it easier for developers to find code examples and learn from them.

This paper proposes Typhon, an approach to automatically recommend relevant code cells in Jupyter notebooks. Typhon tokenizes developers' markdown description cells and looks for the most similar code cells from the database using text similarities such as the BM25 ranking function or CodeBERT, a machine-learning approach. Then, the algorithm computes the similarity distance between the tokenized query and markdown cells to return the most relevant code cells to the developers.

We evaluated the Typhon tool on Jupyter notebooks from Kaggle competitions and found that the approach can recommend code cells with moderate accuracy. The approach and results in this paper can lead to further improvements in code cell recommendations in Jupyter notebooks.
\end{abstract}

\IEEEpeerreviewmaketitle

\section{Introduction}
\label{sec:intro}

Code recommendation has been a great advantage to a lot of software developers for auto-completing the written code or suggesting the next tokens or statements to the developers. 
This function has helped improve developer productivity significantly. It works by taking code the the developer is currently working on and suggesting the complete code snippet back to the user. This reduces the time the developers need to complete their code tremendously.

At present, there are code recommendation tools with two kinds of functionalities: code-to-code recommendation and text-to-code recommendation. A code-to-code recommendation is an approach where the recommendation tools understand the current context of developers' code and recommend the code accordingly. There have been multiple studies of code-to-code recommendation such as Aroma~\cite{Luan2019}, Senatus~\cite{Silavong2022}, Strathcona~\cite{Holmes2005}, and Example Overflow~\cite{Zagalsky2012} using techniques such as clustering and ranking of code snippets, code structure, Jaccard similarity and Minhas-LSH on code, and TF-IDF.
Text-to-code recommendations have developers inquire about the details and behaviors of the code they require from the recommendation tool using natural language text. The example of the existing approaches for text-to-code recommendation include GitHub Copilot\footnote{https://github.com/features/copilot} and Tabnine~\footnote{https://aiterms.net/tabnine/}~\cite{WhatIT}.


Currently, programming is not limited to only building software, but also data analysis. In the data science domain, there are many various kinds of IDE tools that let developers and researchers work on their projects. The most commonly used platform is Jupyter Notebook, which enables a simplified and easy-to-navigate workflow, making Jupyter Notebook one of the most popular options in data Science project development~\cite{Kluyver2016}.


In this paper, we are interested in code recommendations for the Jupyter Notebooks platform by reusing code snippets that are already written by data science developers. We want to study if we can give appropriate code recommendations based on a given markdown text in Jupyter Notebooks. This can help reduce the developer's time by not writing the whole code by themselves or learning from the recommended code snippets and improving further upon them.

To support our motivation, we performed a preliminary experiment of code cell recommendation in Jupyter Notebooks using GitHub Copilot on the content of 5 notebooks from the KGTorrent dataset~\cite{Quaranta2021}. We gave the markdown cell queries and checked the returned Python whether they were relevant to the given markdown cells or not.
The result showed that the GitHub Copilot efficiency offers varied performance with a correctness percentage ranging from a minimum of 0.00 percent to a maximum of 51.16 percent, with an average of 21.83 percent. We investigated the results and found that it is potentially due to the specific environment of the Jupyter Notebook, which has mixes of natural language (markdown text) and code.
In general, Copilot will take a partial comment, which is supposed to be relatively short, and the whole source code file as input and suggest back possible solutions the programmers may need.
Many data science notebooks often have long descriptive text in markdown cells. We found that the Copilot sometimes performs source code suggestions poorly when the given input comment is large, and the whole file context contains a long descriptive text. The given situation often produces non-reusable source code or a comment string related to the input comment instead of the source code.

Therefore, this paper attempts to propose an alternative source code suggestion tool that provides recommended source code from a given descriptive markdown text as input, regardless of the size of the description. Moreover, our proposed technique will not be based on code generation, but on code reuse. 
Our approach leverages the practice of code reuse by recommending code cells in Jupyter notebooks using existing code in a large Jupyter notebook code base in Kaggle, the well-known machine learning and data science platform, using the information from the associated markdown text.
The contribution of this paper is \textit{the similarity matching techniques specifically designed for Jupyter Notebook code cells.} 
Our code-to-markdown cell similarity compares a given markdown cell to other markdown cells and code cells in Kaggle, a large-scale Jupyter Notebooks corpus. 

\section{Background}
\label{sec:background}


\subsection{Jupyter Notebooks}

Jupyter Notebook is a web-based interactive computing and documenting tool that functions in terms of a digital document containing cells of text, code, and computing results proposed by Kluyver et al.~\cite{Kluyver2016}. The concept of the Jupyter Notebook evolves from the interactive shell or REPL (Read-Evaluate-Print-Loop), which is the basis of interactive programming. 
Considering that many academic researchers require to document their research progress, they often need to write computer code to perform computative operations, statistical tests, and run simulations. 
Delivering the composed code in academic papers via human descriptive language is not precise and is unreliable for reproducing the implementation code in that paper. 
Typical academic papers often supply the supplement code in a separate section and are often isolated from the written content. The readers must cross-reference the written content and the relevant code. Consequently, this reduces the readability of the papers and might cause inconsistencies in the code and the prose.
Jupyter Notebook aims to improve readability and increase the document's reproducibility and computative results. 
Therefore, the Notebook is designed to support the scientific research documentation workflow by integrating the text, code, and computative results into a single view. Thus, reducing the effort to cross-reference between code and prose. This Notebook lets authors document their papers via an interactive text and code cells component, enabling later editing and re-executing of the code cells. Hence, increases the reproducibility of the interactive academic documents.
Two major platforms provide hosting services and usage for Jupyter Notebook: Google Colab and Kaggle, both owned by Google, and Deepnote which has free and paid Jupyter Notebook services. 

Kaggle\footnote{https://www.kaggle.com} can be thought of as a large warehouse of open-source Jupyter Notebooks and a playground for data scientists and machine learning engineers. Kaggle is a website that allows users to find and publish data sets and provides a web-based Jupyter Notebooks with a data-science environment.
Kaggle users can work together and participate in competitions to solve data science challenges. Due to the popularity of Kaggle Competitions, most users of Kaggle have participated in competitions to challenge each other. Kaggle also has a ranking system for users, and the rank has four categories, including Competitions, Datasets, Notebooks, and Discussions. The rank of each category will depend on the contribution and the prize of each category. According to a large number of competitions, numerous Jupyter Notebooks have been available on Kaggle as public data. Thus, Kaggle is an excellent place to learn from the works of various users, from beginners to experts. 

\subsection{Code and Text Similarity Measurement}

\subsubsection{Vector Representation Models for Text}

Multiple techniques in the Information Retrieval methods are used to recommend similar text in the collection of documents.

Best Matching 25 (BM25) is a bag-of-words retrieval function, which ranks a set of documents regarding the query terms appearing in each document regardless of the proximity within the document. BM25 computes the similarity between the query terms ($Q$) and the comparison documents $D$ using $\sum_{i=1}^n IDF(q_i)\cdot\frac{f(q_i,D)\cdot(k_1+1)}{f(q_i, D)+k_1\cdot(1-b+b\cdot\frac{fieldLen}{avgFieldLen})}$

where $(q_i)$ is the i\textsuperscript{th} query term in $Q$, $IDF(q_i)$ is the inverse document frequency of the i\textsuperscript{th} query term, which measures how often a term occurs in all documents and penalizes terms that are common, $f(q_i,D)$ is the number of times the ith query term occurs in document $D$, $k_1$ is a variable that limits how much a single query term can affect the score of a document, $b$ is a variable that amplifies the effect of the length of a document relative to the average document length, $fieldLen$ is the length of the field (in terms) of the document, $avgFieldLen$ is the average length of the field (in terms) of all documents.

\subsubsection{Vector Representation Models for Code}
Moreover, recently there have been studies about creating code models, which are machine-learning models that derive vectors (i.e., code embeddings) from code snippets. We discuss some of the recent code models below.

CodeBERT~\cite{Feng2020} is a natural language processing (NLP) pre-trained model developed by Microsoft that can understand both programming language and natural language. The paper presents a hybrid objective function that incorporates the pre-training task of replaced token detection to train the model. CodeBERT can be used for various tasks, such as natural language code search and code documentation generation. The model achieves state-of-the-art performance on two natural language programming language (NL-PL) applications and performs better than previous pre-trained models on NL-PL probing. In summary, CodeBERT is a powerful tool for understanding and generating both programming language and natural language, which can be useful for various applications in software engineering.

UniXcoder~\cite{Guo2022}, also developed by Microsoft, is a unified cross-modal pre-trained model for programming languages, that specializes in code understanding and code generation tasks. UniXcoder is a pre-trained NLP cross-modal model that facilitates the communication between natural language and programming languages for code-related tasks. Additionally, UniXcoder is a pre-trained model in the CodeBERT series. It is designed to understand and generate code from natural language queries or descriptions. UniXcoder is a unified cross-modal model that bridges the gap between natural language and programming language by leveraging advanced machine learning techniques. It is pre-trained on a large corpus of natural language text and programming languages, making it more efficient and effective in code-related tasks.

Additionally, UniXcoder uses a mask attention mechanism with a prefix adapter to control the encode-decode behavior. This allows the model to adjust its output based on the input, improving its accuracy and relevance to the task at hand. Additionally, the model leverages cross-modal content, such as abstract syntax trees and code comments, to enhance code representation. This improves the quality of the code generated by the model and makes it more natural and readable.
UniXcoder is capable of both code-related understanding and generation tasks. This means that it can understand natural language queries or descriptions and generate the corresponding code, or it can take code as input and generate natural language descriptions or comments about the code. This versatility makes it a powerful tool for developers, researchers, and other professionals working with natural language and programming languages. A similarity score between two UniXcoder vectors $A$ and $B$ can be computed using cosine similarity $\cos (\theta ) =   \dfrac {A \cdot B} {\left\| A\right\| \left\| B\right\|}$.

\section{Proposed Methodology: Typhon}
In this paper, we propose a novel methodology, called \textbf{Typhon}, for recommending code cells based on existing code cells from Jupyter notebooks in the Kaggle dataset. 

\subsection{Overview of the Approach}

\begin{figure}[tb]
    \center
\includegraphics[width=0.8\columnwidth]{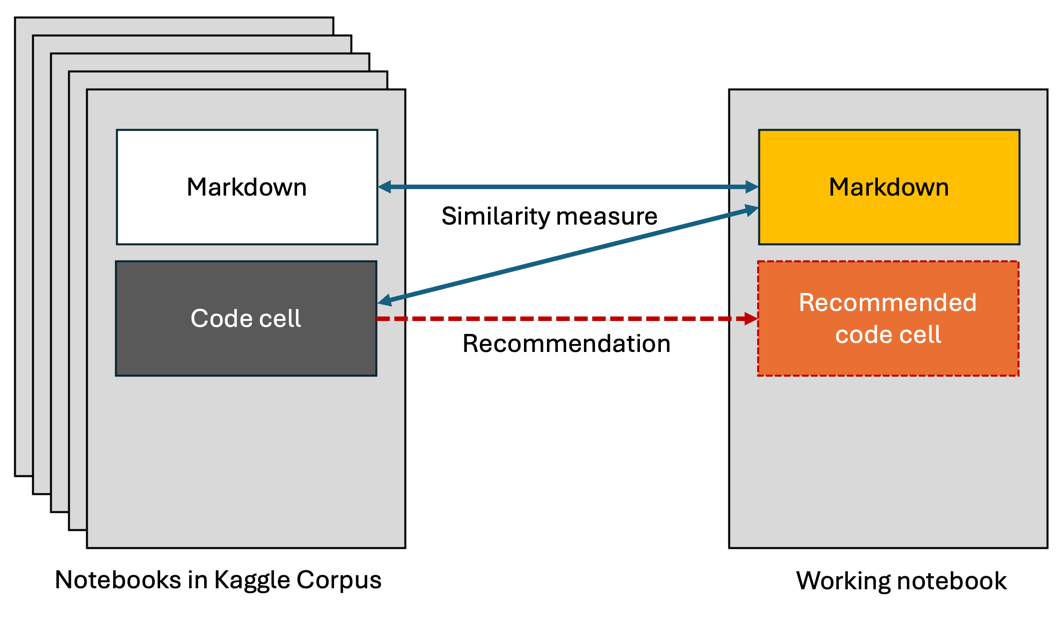}
    \caption{Typhon's code cell recommendation}
    \label{fig:framework-idea}
\end{figure}

As mentioned previously, existing code recommendation tools may not work well with the Jupyter Notebook environment. Where the markdown description cells are varied in terms of length and detail. The recommendation tool often cannot recommend reliable code cells for users because the code and text in the local file's context sometimes provide vague and unclear context.
Therefore, our proposed approach focuses on suggesting code cell(s) based on the similarity of the markdown cell's content to the existing markdown cells or code cells. The concept of the code cell recommendation by the approach is presented in the \autoref{fig:framework-idea}.
The approach will search through the existing corpus of Jupyter notebooks for markdown cells with the most similarity with the input markdown cell. Then, it suggests the code cell with the most similarity to the input markdown cell back.






\subsection{Jupyter Notebook Corpus: KGTorrent}
We leverage KGTorrent, an extensive dataset of computational notebooks with dense metadata retrieved from the Google-owned platform Kaggle, as the corpus of Jupyter notebooks in this paper. The collection process of Jupyter notebooks in KGTorrent covers the time period of November 2015 to October 2020. The data is collected from around 5,598,921 users and 2,910 competitions. KGTorrent has a size of 175 GB and is compressed to the size of 76 GB and consists of 248,761 Jupyter notebooks in the Python programming language. 

\subsection{Markdown-to-Code Cell Recommendation Techniques}

The following are the techniques to locate the most similar code cells according to a given markdown cell that we adopt in this paper.

\subsubsection{BM25}
\begin{figure}[tb]
    \center
    \includegraphics[width=\columnwidth]{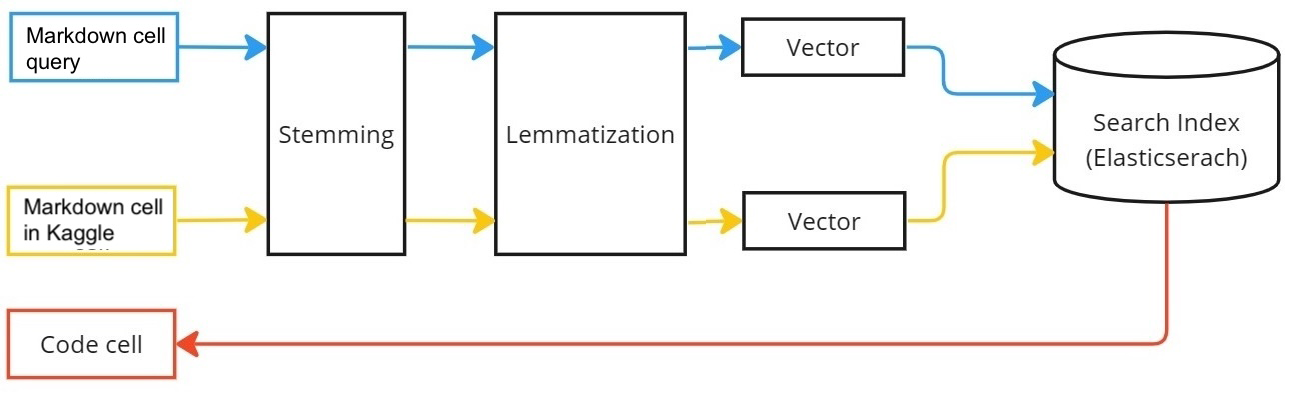}
    \caption{Code cell recommendation using BM25}
    \label{fig:bm25-approach}
\end{figure}

The flow of the BM25 recommendation is depicted in \autoref{fig:bm25-approach}.
The flow starts with the system taking a query markdown cell as input, and then computing the similarity score with markdown cells from Kaggle notebooks in the database using BM25. The more similar the query markdown cell is with the Kaggle markdown cells, the higher the similarity score. After that, the system delivers the code cells corresponding to the matched markdown cells with the highest similarity.
We use Elasticsearch for data indexing, the Python script was used to upload the data that we processed from the raw data of KGTorrent by mapping the markdown cell and code cell next to the markdown cell(s) to be a pair.
Subsequently, we get the set of markdown cells as a query to retrieve comparable markdown pairs from Elasticsearch. We use the default setting of Elasticsearch where BM25 is the similarity and ranking computation method. 

\begin{figure}[tb]
    \center    \includegraphics[width=0.9\columnwidth]{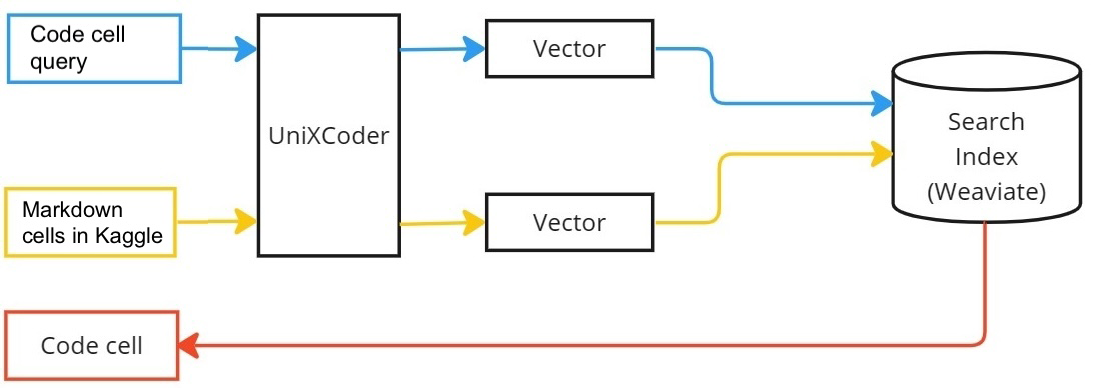}
    \caption{Code cell recommendation using UniXcoder}
    \label{fig:unixcoder-approach}
\end{figure}

\subsubsection{UniXcoder}
The flow of UniXcoder recommendation is as depicted in \autoref{fig:unixcoder-approach}
A query markdown cell is taken into a UniXcoder model to produce an embedding vector representing the input. Similarly, the code cells in Kaggle notebooks are also passed to UniXcoder to create embedding vectors and store them in a database for querying. For UniXcoder, we search for code cells by computing the similarity of the code cells directly with the markdown cell because UniXcoder is a cross-model. Typhon utilizes the Weaviate\footnote{https://weaviate.io}, a vector search engine database for accommodating the vectorizing process and storing a collection of code datasets. Both the similarity-finding method and returning such results happen within Weaviate. Then, our script returns the recommended code cells. We use the default similarity setting of Weaviate which is cosine similarity.

\section{Evaluation}
We evaluate Typhon in two ways: recommendations of code cells containing Matplotlib plots and a sanity check. 

\begin{table*}[tb]
\caption{Matplotlib visualization types and the associated query terms}
\label{table:query-template}
\centering
\begin{tabular}{p{3cm}p{4cm}l}
\toprule
Plot type & Sub plot type & Query term \\
\midrule
Basic &	Scatter &	plot data using scatter visualization \\
	& Bar &	plot data using bar visualization \\
	& Stem &	plot data using stem visualization \\
	& Step &	plot data using step visualization \\
	& Fill\_between &	plot data using fill\_between visualization \\
	& Stackplot &	plot data using stackplot visualization \\
 \midrule
Plots of Arrays and Fields &	Imshow &	plot data using imshow visualization \\
	& Pcolormesh &	plot data using pcolormesh visualization \\
	& Contour &	plot data using contour visualization \\
	& Contourf &	plot data using contourf visualization \\
	& Barbs &	plot data using barbs visualization \\
	& Quiver &	plot data using quiver visualization \\
	& Streamplot &	plot data using streamplot visualization \\
 \midrule
Statistics Plots & Hist &	plot data using hist visualization \\
	& Boxplot &	plot data using boxplot visualization \\
	& Errorbar &	plot data using errorbar visualization \\
	& Violinplot &	plot data using violinplot visualization \\
	& Eventplot &	plot data using eventplot visualization \\
	& Hist2d &	plot data using hist2d visualization \\
	& Hexbin &	plot data using hexbin visualization \\
	& Pie &	plot data using pie visualization \\
 \midrule
Unstructured Coordinates &	Tricontour &	plot data using tricontour visualization \\
	& Tricontourf &	plot data using tricontourf visualization \\
	& Tripcolor &	plot data using tripcolor visualization \\
	& Triplot &	plot data using triplot visualization \\
 \midrule
3D & 3D Scatterplot &	plot data using 3D scatterplot visualization \\
	& 3D Surface &	plot data using 3D surface visualization \\
	& Triangular 3D Surface &	plot data using triangular 3D surface visualization \\
	& 3D Voxel , Volumetric Plot & plot data using 3D voxel , volumetric plot visualization \\
	& 3D Wireframe Plot &	plot data using 3D wireframe plot visualization \\
\bottomrule
\end{tabular}
\end{table*}
    
 
\subsection{Recommendations of Code Cells Containing Matplotlib Plots}

For the recommendation of the code cells containing Matplotlib plots, we first studied the available plot types from the Matplotlib website. This is followed by creating a template query text as a representative of the markdown cell for all the plot types, which is shown in \autoref{table:query-template}. The next step is to start querying each suggested code and determining whether the suggested code is related to the query chart type.

Before starting the evaluation process, the codebase data of both BM25 and UniXcoder were prepared and organized into three groups based on their corresponding ranks of the notebooks' authors in Kaggle. 
The numbers of code-markdown pairs in each rank group were as follows: 2,670 pairs for Rank Grandmaster, 4,098 pairs for Rank Master, and 10,855 pairs for Rank Expert. 
These numbers are approximate because some pairs were discarded during the indexing operation in the Machine Learning model.
The data in the codebase was indexed from the KGTorrent dataset by mapping raw markdown-code cells and filtering for only code-markdown pairs related to plotting and charts to reduce the scope of the evaluation.

The third author who has a background in Data Science is tasked with comparing the suggested code with the query chart type.
If the suggested code is related to the query chart type, then the recommendation is marked as correct. On the other hand, if it is not related, the recommendation is marked as incorrect. This process is carried out for all the plot types on the Matplotlib website.
Finally, after completing querying recommendations for all plot types, the results are used for evaluating the accuracy of recommendations.  



The result from the evaluation is shown in \autoref{table:final-exp-result}, where it included 14 chart types collected from the Matplotlib library \footnote{https://matplotlib.org/stable/plot\_types/index}websites. 
\autoref{table:final-exp-result} contains the Plot Type column, 3 rank-level columns; Grand Master, Master, and Expert, and 2 sub-types for model columns; UniXcoder and BM25. At the Grand Master rank, UniXcoder has the capability to suggest 6 code snippets that are related to the charts and 3 code snippets for BM25. As for the Master rank, UniXcoder can recommend 7 related code snippets, and the BM25 approach can recommend for 4 recommendations. Finally, for the Expert rank, UniXcoder can recommend 9 related code snippets and 7 related code snippets for the BM25 approach. 

\begin{table*}[tb]
\caption{Typhon's first recommendation contains the code cells related to the given visualization type.}
\label{table:final-exp-result}
\centering
\begin{tabular}{lcccccc}
\toprule
& \multicolumn{2}{c}{Grand Master} & \multicolumn{2}{c}{Master} & \multicolumn{2}{c}{Expert} \\
\cmidrule{2-7} 
Plot Type & UniXcoder & BM25 & UniXcoder & BM25 & UniXcoder & BM25 \\
\midrule
scatter & \checkmark & \checkmark & \checkmark & \checkmark & \checkmark & \checkmark \\
bar & \checkmark & & \checkmark & \checkmark & \checkmark & \checkmark \\
step & \checkmark & & & & \checkmark & \checkmark \\
imshow & \checkmark & & \checkmark & & & \checkmark \\
contour & &	& & & \checkmark & \\
hist & & & & & & \checkmark \\
boxplot & & \checkmark & & & \checkmark & \checkmark \\
errorbar & & & \checkmark & & & \\
violinplot & & & & \checkmark & \checkmark & \\
pie & \checkmark & & \checkmark & \checkmark & & \\
tripcolor & \checkmark & & & & & \\
3d scatterplot & & \checkmark & \checkmark & & \checkmark & \checkmark \\
3d surface & & & \checkmark & & \checkmark & \\
triangular 3d surface & & & & & \checkmark & \\
\midrule
Total Correct & 6 & 3 & 7 & 4 & 9 & 7 \\
\bottomrule
\end{tabular}
\end{table*}

\subsection{Sanity check}
\begin{table*}[tb]
\caption{Sanity check result}
\label{table:sanity}
\centering
\begin{tabular}{llrrr}
\toprule
\textbf{Rank} & \textbf{Type} & \textbf{Total Items} & \textbf{Total Correct}& \textbf{Total Correct (\%)} \\
\midrule
\multirow{3}{3cm}{Grand Master} & UniXCoder & \multirow{3}{*}{2,517} & 254& 10.01\\
                                & BM25 & & 2,399 & 95.31 \\
 & BM25 + stemming and lemmatization & & 2,132 & 84.72\\
\midrule
\multirow{3}{3cm}{Master}  & UniXCoder & \multirow{3}{*}{3,744} & 377 & 10.07\\
                           & BM25 & & 3,391 & 90.57 \\
 & BM25 + stemming and lemmatization & & 3,007 & 80.32\\
\midrule
\multirow{3}{3cm}{Expert}  & UniXCoder & \multirow{3}{*}{9,553} & 605 & 6.33 \\
                           & BM25 & & 8,644 & 90.48 \\
 & BM25 + stemming and lemmatization & & 7,193 & 75.30 \\
\bottomrule
\end{tabular}
\end{table*}

We also performed a sanity-checking test aiming to make sure that Typhon can provide recommendation functionality, at the very least, as correct as it should be.
The steps are in the following order. First, each code-markdown pair in the code base is used to query for recommendations from the same Kaggle dataset. Then, those recommended code cells for UniXCoder, BM25, and BM25 with the pre-processed query using stemming and lemmatization are compared with the original code from the query code-markdown pair.

During the sanity check, the data used for both the databases and queries were the same, therefore they were identical. We scope down the data in this sanity check. Again, this data consisted of code-markdown pairs specifically related to plotting and charts. 
There were 2,670 code-markdown pairs in Rank Grandmaster, 4,098 pairs in Rank Master, and 10,855 pairs in Rank Expert.
The markdown in each code-markdown pair was used as the query.

Our assumption expects that the recommendation result should return the exact copy of the original code in the query code-markdown pair.
The sanity-check result is shown in the \autoref{table:sanity}. 
The result comprises 3 primary ranks of the notebook authors in Kaggle:~Grand Master, Master, and Expert. 
Each level contains 3 types of the recommendation model approach available for use in the Typhon tool, with the Total Correct, Total Correct (\%), and Total Items columns, explaining the total number of correct recommendations, the total number of items in the codebase, and the percentage of correct items.

We analyze the result according to the rank of the Kaggle notebook's authors.

\begin{enumerate}
    \item In the Grand Master rank, UniXCoder has the least correct recommendation, with 254 items or 10.01 percent. The correct recommendation is the BM25 model, with 2,399 items, or 95.31 percent. In comparison, BM25 with stemming and lemmatization has the correct recommendation of 2,132 items or 84.72 percent.
    \item In the Master rank, the ranking is the same; UniXcoder has the least correct recommendation items, the BM25 model is the highest, and BM25 has a pre-processed query in the middle. Where the UniXCoder has 377 items or 10.07 percent, BM25 has 3391 items or 90.57 percent, and BM25 with pre-processed query has 3,007 items or 80.32 percent.
    \item In the Expert rank, the rank is also the same. UniXCoder has the least correct recommendation items, the BM25 model the highest, and BM25 with the pre-processed query in the middle. UniXcoder has 605 items or 6.33 percent, Elasticsearch has 8,644 items or 90.48 percent, and Elasticsearch with pre-processed query has 7,193 items or 75.30 percent.
\end{enumerate}

The low recommendation score in the UniXCoder could be because UniXcoder is a cross-modal model that contains only code cell data in the codebase and returns recommendations by processing a natural language query with the code in the codebase, unlike Elasticsearch, which has pairs of markdown cells and code cells for cross-referencing.
On the contrary, the score in BM25 and BM25 with pre-processed queries using stemming and lemmatization is significantly higher than BM25 and UniXcoder. BM25 suggests the code by comparing the natural language in the query with the natural language in the markdown in the objects stored in the Elasticsearch codebase and pre-processing helps increase the accuracy.

We also performed a further investigation of the BM25 recommendations that were wrong. The authors found that there were several issues that made the BM25 approach recommend different codes, even though the associated markdown was used for querying already existed in the codebase. The following section discusses the topics found in the problem.

\begin{enumerate}
    \item General used words:~The query markdowns are not scoped into any particular topic or context. These markdowns are short in length and convey very broad meaning. Examples of these markdowns are ``Other'' and ``Worldwide''.
    \item Few keywords or short phrase markdown:~The query markdown consists of cases where the markdown used for querying contains only a few keywords or short phrases. While the markdowns in the recommendation are often longer and contain more of the keywords from the original markdowns.
    \item Reused and duplicated markdowns:~There are many cases in which the author reuses and replicates the same markdown multiple times, both in the same notebook and in different notebooks. There are also cases where different authors use the same markdown in their own notebooks, which contain different codes. Often, the markdown in such cases is short or has few keywords.
    \item General inaccurate recommendation:~Various common inaccurate results can be noticed from the results, in which there were cases where the query markdown and recommendation markdown shared the same keywords that are often unique among notebooks, or significantly common that can be found in any notebook.
\end{enumerate}

The investigation clearly shows that clear and detailed instructions in the markdown can directly improve code recommendation quality and improve the likelihood of generating relevant and accurate suggestions.

However, while the BM25 recommendation approach provides valuable insights and potential code recommendations, it is important to note that the algorithm may not always capture the specific context or subtle requirement in the markdown query. Human experts are still required to manually intervene to verify and fine-tune the recommended code to ensure that actually relevant recommended codes are selected from the recommendation.
To enhance the BM25 approach and address the observed issues, further research and refinement of the recommendation algorithm, as well as incorporating additional contextual information from the markdown, could potentially lead to more accurate and reliable code suggestions. We leave this as future work.

\section{Related Work} 
According to the study of Ritta et al.~\cite{Ritta2022}, they investigated if code reuse occurs in Jupyter Notebooks of Kaggle Competition. They selected three competitors that have high-level skills in writing and programming in Jupyter Notebooks. After that, they extracted all notebooks from all the competitions of three competitors. The code cells extracted from each competitor's latest competitions will be used as queries to find the reused code cells via NCDSearch\footnote{https://github.com/takashi-ishio/NCDSearch}.
NCDSearch is a grep-like tool. One command of a Unix-like operating system named grep is used for searching plain text from lines that match a regular expression in files. Thus, NCDSearch is created from the grep concept to find similar source code fragments in specific files.
Any code cell with a similarity distance from 0 to 0.5 was counted as reused code. They also classified and found the common type of reused code to see which type is the most code that is reused. They discovered that competitors might have different most reused code types, but the most common type that everyone reused is Import Package, which is code used for importing Python's packages.
Additionally, the authors suggest that their work has the potential to be fundamental for the automatic code-recommending tool because code recommendation can come from reusing the existing codes from open-source projects.

\textbf{Aroma} is a source code recommendation tool that is implemented with a code-to-code search concept from Luan et al.~\cite{Luan2019}. The core idea of Aroma is recommending any possible complete and working or extended code snippets, having the input as partially written source code. Aroma's purposes are to address the extensibility opportunity of the existing code snippet, such as error handling parts and setup of the code, and comparison with the common pattern other developers have made. It takes a code snippet as input, and then Aroma will index the given corpus of code and retrieve a small set of code snippets, where the retrieved code snippet will approximately contain the input code snippet. Then Aroma will take these retrieved snippets through a series of pruning, ranking, and clustering. After clustering, Aroma filters these clustered snippets with intersecting operations to derive a set of the most common snippets among the retrieved snippets. These intersected snippets are recommended to the user as a set of alternative extended code snippets. Although Aroma shares the same concept of recommending source code, the input of Aroma is taken as a snippet of code, whereas our paper focuses on the input as markdown cells.

\textbf{Senatus} is a code-to-code recommendation engine for developers that aims to enhance their productivity, efficiency, and the reliability and consistency of the code. The team builds Senatus by following the definition given by Luan et al.~\cite{Luan2019}. The authors propose a new method of code recommendation by indexing the code repository with Minhash-Locality Sensitive Hashing or Minhas-LSH. Minhash is generally used for maintaining the Jaccard similarity of sets by representing those sets as a lower dimensional similarity and preserving features of their vectors. This technique provides much better efficiency at the query time. Minhash can be applied in many applications such as web search, graph sampling, earthquake detection, and efficient clustering of a very large collection of DNA sequences. However, by using Minhash in the code-to-code recommendations, they provide poor retrieval quality because of the length of skewness in the code snippet. Senatus fixes this issue by operating directly on the abstract syntax tree (AST) structure representation of the code which decreases the time to query and the effectiveness of the retrieval.


\textbf{Strathcona:~}
Another benefit of code recommendation is provided by the Strathcona Example Recommendation Tool from Holmes et al.~\cite{Holmes2005}, a tool that helps developers understand and use API. Incomplete or outdated documentation about API usage causes problems for developers, so the tool aims to reduce the gap between source code and documentation. In the case of understanding API usage, developers must spend a lot of time manually searching the relevant source code showing how to use the API. With the example recommendation tool, developers can write code and get recommended examples for API usage. The authors of the work intended to make the developers use less effort to get examples. The code written by developers will be a query for searching. Then the tool will locate structurally relevant examples by comparing the context of the structure between the query and the existing repository. Moreover, after the tool finds the relevant examples and returns them, developers are able to navigate the example to inspect it in more detail as desired. 
Returned examples can be navigated visually and textually.

\textbf{Example Overflow:~} Using the idea of social media or wisdom of crowd recommendations, Zagalsky et al.~\cite{Zagalsky2012} propose Example Overflow, a code search, and recommendation tool. Example Overflow is a generic recommendation system that utilizes the code snippet and textual data on Stack Overflow. The approach's core idea is to retrieve the relevant code snippets based on human evaluation. Other existing search tools such as Strathcona, Aroma, and Senatus recommend relevant code snippets using the machine processing power to compute the rank and similarity. On the other hand, Example Overflow leverages the crowd wisdom from answers to questions from the Stack Overflow website. Intuitively, many programmers of any experience often learn and code by following examples of code snippets in the tool documentation or code examples on the Internet. Therefore, Example Overflow offers two core practices by the system. The first practice is comparing multiple examples simultaneously without switching developing context, either by switching the browser tabs or searching for code examples. The system aims to serve this key point by presenting the top five most relevant results, all in the same view. In the second practice, the system takes minimal context switching, according to the first practice, where the programmers are presented with the top five results. 

There are also some of the existing work in text-to-code recommendation tools and techniques as discussed below.

\textbf{CodeSearchNet} is a collaboration between GitHub and the Deep Program Understanding group at Microsoft Research~\cite{Husain2019}. It is a dataset and benchmark for code retrieval techniques. The goal of CodeSearchNet is to allow developers and researchers to develop and improve code retrieval techniques which are essential for improving software development productivity and quality.
CodeSearchNet dataset is collected from publicly available open-source repositories on GitHub. The dataset includes over 6 million methods consisting of GO, Java, JavaScript, Python, PHP, and Ruby programming languages. Of these, 2 million methods have associated documentation. This results in a set of pairs between the method of code and its documentation.
CodeSearchNet also provides a benchmark framework to measure and compare the performance of different code retrieval techniques. Its purpose is to encourage researchers and practitioners to further study semantic code search tasks, which involve retrieving relevant code given a natural language query. The process is generally as follows: 99 queries from the natural language with its corresponding possible programming language results in each considered language (Go, Java, JavaScript, PHP, Python, and Ruby). Then, these query/results pairs are labeled by human experts to indicate the relevance of the result and the query. Finally, the experts annotate the result for relevance on a zero to three scale. The framework supports exact match retrieval, semantic retrieval, and cross-lingual retrieval~\cite{Husain2019}.
On the leaderboard of the CodeSearchNet benchmark 4 out of the top 5 results use the NeuralBoW (Neural Bag-of-Words). The basic idea behind NeuralBoW is to first represent each word in a document using a one-hot encoding to create a vector of zeros except for a single 1 in the position corresponding to the index of the word in a pre-defined vocabulary. These vectors are then averaged together to produce a fixed-length vector representation of the entire document. This document vector can then be used as input to a neural network classifier to predict the document's class label. 


We compare our Typhon approach to existing code-to-code and text-to-code recommendation techniques as shown in Table~\ref{table:comparisons}.
The seven techniques, including Typhon, have different ways of recommending code. Copilot and Tabnine use machine-learning techniques to generate new code from scratch. The other five techniques; Aroma, Strathcona, Senatus, ExampleOverflow, and Typhon; are based on searching from existing code bases by using the other code snippets or search keywords as queries. Aroma recommends clustering and ranking code snippets categories, Strathcona recommends from the similarity between the user's IDE structural details and code structural detail in repositories, Senatus recommends from the similarity of input description and database using Jaccard similarity, Example Overflow recommends from fine-tuned TF-IDF weight algorithm, and Typhon suggests code cells using BM25 and UniXcoder.

\begin{table}[tb]
\caption{Comparison of Typhon to Previous Work}
\label{table:comparisons}
\begin{tabular}{lp{6cm}}
\toprule
Name & Approach \\
\midrule
Typhon & Search and recommend code cells based on the similarity of markdown text cells using BM25 or Machine Learning. \\
Copilot & Generate code cells using the OpenAI Codex model trained on large open-source projects in GitHub \\
Tabnine & Generate code snippets based on AI-based proprietary algorithm \\
Aroma & Search and recommend code snippets based on the clustered and ranked code snippet categories\\
Strathcona & Search and recommend code snippets based on the user's IDE structural details with relevant to structural details in code repositories\\
Senatus & Search and recommend code snippets based on Jaccard similarity using Minhash-LSH \\
ExampleOverflow & Search and recommend code snippets by keyword search using fine-tuned TF-IDF weight\\
\bottomrule
\end{tabular}
\end{table}


 \section{Threats to Validity}
\textit{Internal Validity:}~The number of code-markdown pairs in the codebase is inconsistent due to the limited amount of available data in the dataset and filtering process. 
Where the Expert rank has the most data, with approximately 10,800 pairs of code and markdown; the Master rank with approximately 4,090 pairs; and the Grand Master rank with approximately 2,660 pairs. 
Therefore, the present imbalance could explain the noticeably slightly higher accuracy in Master and Expert, where they can potentially recommend a more variety of code due to more available data.
Furthermore, the charts presented in \autoref{table:final-exp-result} show that there are some charts that rarely appeared in the recommendation results. 
Additionally, considering the nature of the dataset, which is extracted from Kaggle competition, there is a high possibility that the codebase contains some group of visualizations significantly more than other groups of visualization. 
Therefore, it explains why some charts appear more often than others appear in the recommendation. This may affect the validity of the results.

\textit{External Validity:}~The proposed approach works with only the data from Kaggle and may not be generalized to Jupyter notebooks on other platforms. Moreover, the similarity computations are based only on BM25 and UniXcoder on English text in the markdown and Python code which may not be generalized to other techniques and languages.

 \section{Conclusion and Future Work}
Code recommendation tools have become a big part of many software developers. 
Tools like GitHub Copilot and ChatGPT have become increasingly favored both in and outside the industry. 
Our proposed approach, Typhon, recommends similar code cells in Jupyter notebooks using BM25 (Information Retrieval Approach) and UniXcoder (Machine Learning Approach). 
Typhon is designed to improve the workflow of those in the Data Science field, primarily on Jupyer Notebook.

The evaluation results show that Typhon can recommend code cells involving visualization using Matplotlib with moderate accuracy. 
UniXcoder can recommend more correct code cells than BM25. Nonetheless, our sanity check shows that UniXcoder has lower precision than BM25. 

In future work, we plan to compare the performance of Typhon to other approaches such as Copilot. M
oreover, we plan to improve the accuracy of the recommendation by increasing the number of markdown cells that are used to query for similar code cells and study its effect on the recommendation accuracy. We also plan to explore other code embedding techniques besides UniXcoder.

\section*{Acknowledgement}
This work was financially supported by the Office of the Permanent Secretary, Ministry of
Higher Education, Science, Research and Innovation (OPS MHESI) Grant No. RGNS 64-143, and the Faculty of Information and Communication Technology, Mahidol University, Thailand.
 
\bibliographystyle{IEEEtran}
\bibliography{references}

\end{document}